\documentclass[12pt,a4paper,notcite,notref]{article}
\usepackage{amsmath}
\usepackage{cite}
\title{Algebraic entropy for differential-delay equations} 

\author{C-M. Viallet \\  Sorbonne Universit\'es, UPMC Univ Paris 06 \\
 Centre National de la Recherche Scientifique \\
 UMR 7589, LPTHE, F-75005, Paris, France }

\begin{document}
\maketitle

\begin{abstract}
We extend the definition of algebraic entropy to a class of
differential-delay equations. The vanishing of the entropy, as a
structural property of an equation, signals its integrability. We
suggest a simple way to produce differential-delay equations with
vanishing entropy from known integrable differential-difference
equations.
\end{abstract}

\section{Introduction}

Algebraic entropy was introduced as a measure of complexity and a test
of integrability of maps~\cite{BeVi99}, an elaboration on the results
of~\cite{Ve92,FaVi93}. It associates to any (bi)rational map over
projective space an invariant quantity characterising its complexity.
The drop of complexity signalled by the vanishing of algebraic entropy
is the signature of integrability. The entropy usually happens to be
calculable exactly, and enjoys remarkable properties, making it an
object of interest {\em per se}. It has for example been conjectured
to always be the logarithm of an algebraic integer~\cite{BeVi99} and
such algebraicity property of entropies are currently under study, see
for example~\cite{DiHaKaKo14}.

Initially devised for systems with a finite number of degrees of
freedom, the concept has been extended to systems with an infinite
number of degrees of freedom, by going from maps (ordinary difference
equations) to lattice equations (partial difference
equations)~\cite{TrGrRa01,Vi06}, and further to systems mixing
differential and difference equations~\cite{DeVi12}. We present here
an extension to a class of  differential-delay equations.

The question of the integrability of this type of equation has already
been addressed in the
literature~\cite{QuCaSa92,GrRaMo93}~\cite{Jo08,JoSp09}, leading in
particular to the concept of delay Painlev\'e equations, one source of
such equations being the symmetry reductions of integrable lattice
equations, or semi discrete equations.

Of course we have gone far away from the original definition of
integrability given by Liouville for hamiltonian systems, but we wish
to call integrable any system with vanishing algebraic entropy.

The equations we consider are recurrences of order $k$ of the form
\begin{eqnarray}
\label{delaydiff}
u(t+1) = R( u(t), \;  u(t-1), \; \dots,\;  u(t-k+1))
\end{eqnarray}
with $R$ a rational function of $ u(t),\; u(t-1),\; \dots, \;
u(t-k+2)$, and their derivatives, and homographic in $u(t-k+1)$,
possibly depending explicitly on $t$ in the non autonomous case.  The
homographic nature of (\ref{delaydiff}) with respect to $u(t-k+1)$
allows to define a {\em birational} map from $ u(t), u(t-1), \dots,
u(t-k+1)$ and their derivatives to $ u(t+1), \; u(t), \; \dots,\;
u(t-k+2)$ and their derivatives.

The setting is thus very similar to the one of~\cite{DeVi12}. The
crucial difference is that the derivative and the shift act on the
same variable $t$. We will nevertheless consider the values of the
unknown function $u$ at the successive points $\dots, t-1, t, t+1,
\dots$ as {\em independent} members of a sequence as was done
in~\cite{GrRaMo93,GrRa97b}, and write
\begin{eqnarray*}
u_k(t) = u(t+k).
\end{eqnarray*}

{\bf Caveat}: Accumulating infinitely many derivative of some $u_n$ amounts,
if the Taylor expansion of $u_n$ converges, to giving more than one
element of the sequence of $u$'s. We will come back to this point
later.

The definition of the entropy then reproduces the one given
in~\cite{DeVi12}.  We use $k+1$ homogeneous coordinates to manipulate
only polynomial differential expressions. Defining

\begin{eqnarray*}
\left[
\begin{matrix}
u_n &= &x_1/x_{k+1} \\
u_{n-1} & =&  x_2/x_{k+1} \\
& \dots & \nonumber \\
u_{n-k+1} &=& x_k/x_{k+1} 
\end{matrix}
\right]
\qquad  \mbox { and } \qquad
\left[ 
\begin{matrix}
u_{n+1} &= &y_1/y_{k+1} \\
u_{n} & =&  y_2/y_{k+1} \\
& \dots & \\
u_{n-k+2} &=& y_k/y_{k+1}
\end{matrix}
\right].
\end{eqnarray*}

Equation (\ref{delaydiff}) can be rewritten as a differential
polynomial map
\begin{eqnarray}
\varphi: [x_1, \dots, x_{k+1}] \longrightarrow  [y_1, \dots, y_{k+1}]
\end{eqnarray}

The inverse $\psi$ is also a differential polynomial map.  One may
define the quantity $\kappa_\varphi$ and $\kappa_\psi$, which are
useful for the singularity analysis. Recall that
$\kappa_\varphi$ (resp.  $\kappa_\psi$) are the multipliers appearing
when composing $\varphi$ and $\psi$ into $\psi \cdot \varphi$ (resp
$\varphi \cdot \psi$), both these products being the identity
operator (see~\cite{BeVi99}).

The successive iterates produce a sequence of polynomials in the
$u_k$'s and their derivatives. Attributing a weight $1$ to all $u_k$
and their derivatives yields the sequence $\{ w_n\}$ of the weights of
the iterates (Leibnitz rule ensures that all components have the same
weight at all stages). Moreover the straightforward property $ w_{l+m}
\leq w_l \cdot w_m$ ensures the existence of the limit
\begin{eqnarray}
\label{entropy}
\epsilon = \lim_{m \rightarrow \infty} \frac{1}{m} \log( w_m).
\end{eqnarray}
the {\em algebraic entropy}.

What is remarkable is that the asymptotic behaviour of the sequence of
weights is entirely determined by local features: finite subsequences
contain enough information to extract the entropy.  This shows itself
in most cases by the existence of a finite recurrence relation on the
weights (see~\cite{HaPr05} for a discussion). This is best seen when
one can fit the generating function of the sequence by a rational
fraction\footnote{The fact that $\epsilon$ is the log of an algebraic
  integer appears there}.

This property has two merits:

- it allows to calculate the entropy efficiently, by exempting us from
the calculation of the entire sequence $\{ w_n \}$

- it also allows to consider $u(t)$ and $u(t+k)$ as independent
objects, and thus to use the same method of calculation as
in~\cite{DeVi12}. Since we only use a finite number of iterates, we
will use derivatives only up to a finite order. We then avoid the risk
of a mismatch between the value of $u$ at some point and the value
given by a Taylor expansion from some other point.

\section{An explicit example}

We will  apply this scheme  to the following equation~\cite{QuCaSa92}:
\begin{eqnarray}
\label{qcs}
a \; u(t) - b \; \dot{u}(t) = u(t) \; \left[  u(t+1) - u(t-1) \right] 
\end{eqnarray}
where $\dot{u} $ means time derivative.

The choice is motivated by multiple reasons:
\begin{itemize}
\item Equation (\ref{qcs}) was obtained in~\cite{QuCaSa92} by a non
  trivial reduction of a semi-discrete equation~\cite{Ma74} and should
  inherit the integrability properties of the latter. The existence of
  an induced Lax pair is one.
\item It has rational solutions for specific values of the parameters.
\item One  continuous limit  is Painlev\'e I equation.
\item The Nevanlinna theory approach~\cite{AbHaHe00,HaKo05,HaKo06} was
  successfully applied to this equation~\cite{HaKo14}.
\end{itemize}

Equation (\ref{qcs}) defines a recurrence of order 2 translating into the
map $\varphi$ and its inverse $\psi$.
\begin{eqnarray}
\label{phi}
\varphi: [x_1,x_2,x_3] &\longrightarrow & [ a \, x_1 x_3 - b\, ( x_1' x_3 - x_1 x_3') + x_1 x_2, \; x_1^2,\;  x_1 x_3] \\
\label{psi}
\psi : [x_1,x_2,x_3] & \longrightarrow &[ x_2^2, \; - a\, x_2 x_3 + b\, ( x_2' x_3 - x_2 x_3') + x_1 x_2 , \; x_2 x_3 ]
\end{eqnarray}

were prime ($'$) means derivative.

These birational maps act on the algebra generated by $x_1, x_1', x_1'',
\dots , x_3, x_3',x_3'', \dots$. 

The multipliers $\kappa_\varphi$ and $\kappa_\psi$ are just
\begin{eqnarray*}
\kappa_\varphi ([x_1,x_2,x_3])= x_1^3 , \qquad \kappa_\psi([x_1,x_2,x_3])= x_2^3.
\end{eqnarray*}

The sequence of weights  of the iterates is 
\begin{eqnarray}
\label{sequence}
\{ w_n \} = 1, 2, 4, 8, 13, 20, 28, 38, 49, 62, 76, \dots 
\end{eqnarray}

The generating function of this sequence is fitted by the rational
fraction
\begin{eqnarray}
\label{gener}
g(s) = \sum_n d_n s^n = \frac{1+2s^3}{(1+s) ( 1-s)^3},
\end{eqnarray}
showing quadratic growth of the weights, a standard  signature of
integrability.

Notice that in order to produce the sequence~(\ref{sequence}), we use
arbitrary initial conditions, i.e.  $ x_1,x_2$ and $x_3$ which are
{\em independent} functions of $t$. Recall also that what matters
in~(\ref{gener}) is the location of its poles and the order of the
pole at $s=1$.

We may perform a finer analysis of the factorisation process leading
to this sequence by keeping track of the factors which are removed
from the homogeneous coordinates.

Starting from 
\begin{eqnarray*}
U_0 = [\; A_0, B_0, C_0 \;],
\end{eqnarray*}

we get the sequence
\begin{eqnarray*}
U_1 & = & [\; A_1, \; A_0^2, \; A_0 C_0 \; ] \\
U_2 & = & [ \; A_2  , \; A_1^2   , \; A_0 A_1 C_0 \;  ] \\
U_3 & = & [\;  A_0^2 A_3 ,\;  A_2^2  , \; A_0  A_1 A_2 C_0\;  ] \\
U_4 & = & [\;  A_1^2 A_4,  \;  A_0 A_3^2, \; A_1 A_2 A_3 C_0 \;  ] \\
\dots \\
U_k & = & [\;  A_{k-3}^2 \; A_k,  \;  A_{k-4} \; A_{k-1}^2, \; A_{k-3} \; A_{k-2} A_{k-1} \; C_0 \; ] \\
 \end{eqnarray*}

where the $ A_k$'s are  polynomials in the  initial conditions $U_0$ and their derivatives.

We see that the factorisation pattern stabilises after the fourth iterate,
 which is actually the first order at which a non trivial
factorisation happens, and we will suppose it remains stable.

Denote by $X_k$ the first component of $U_k$. We get the image
$U_{k+1}$ of $U_k$, in two steps. First evaluate $ \varphi( U_k)$
using the homogeneous expression~(\ref{phi}). Any common factor to the
three components is then removed. Call $ f_{k+1}$ this factor. The
previous sequence indicates that beyond the fourth iterate:
\begin{eqnarray*}
f_n \cdot f_{n-3}^2 = X_{n-4}^3.
\end{eqnarray*}
This implies the finite recurrence relation
\begin{eqnarray*}
w_{n+4} - 2 \,  w_{n+3} + 2 \, w_{n+1} - w_n =0.
\end{eqnarray*}
The characteristic polynomial of this recurrence relation is
\begin{eqnarray*}
s^4 - 2 \, s^3 +2 \,  s - 1 = (s+1) \; ( s-1)^3,
\end{eqnarray*}
where we recover the denominator of the generating
function~(\ref{gener}), and the quadratic growth of the weights:
\begin{eqnarray*}
w_n = \frac {1}{8} \; \bigg(  6 \; n^2 +9 - (-1)^n \bigg).
\end{eqnarray*}

{\bf Remark}: The fact that factors of increasing degree appear when
we proceed with the iteration means that the components of $U_k$
verify {\em relations linking consecutive points}, as 

\begin{eqnarray*}
  A_{k} \; A_{k+3} - b \; A_{k+1}' \; A_{k+2} \; C_0 = P \; A_{k+1},
\end{eqnarray*}
for some $P$.  This factorisation property yields
\begin{eqnarray*}
f_k = A_{k-4}^3.
\end{eqnarray*}

\section{Departing from integrability}

It is interesting to see how the picture changes if we deviate from 
(\ref{qcs}), taking for example
\begin{eqnarray}
\label{qcs_devx}
a \; u(t) - b \; \dot{u}(t) = u(t) \; \left[ u(t+1) - \lambda \; u(t-1) \right]
\end{eqnarray}
For generic $\lambda$ we get the sequence
\begin{eqnarray}
\{ w_n \} = 1, 2, 4, 8, 16, 32, 64,  \dots 
\end{eqnarray}
that is to say no factorisation, and algebraic entropy $ \epsilon=
\log(2)$.  

The proof is straightforward: the multiplier $\kappa_\varphi$ is slightly modified.
\begin{eqnarray*}
\kappa_\varphi( [x_1, x_2, x_3]) =  \lambda \; x_1^3.
\end{eqnarray*}
The successive images of the surface $\kappa_\varphi=0$ are the points
$ [-1, 0, 0]$, $[-1, 1, 0]$, $ [1 - \lambda, 1, 0]$, $ [1+\lambda,
  1-\lambda, 0]$, $[(1-\lambda)\lambda, 1+\lambda, 0]$, $[1+\lambda,
  1- \lambda, 0]$ and so on. If $\lambda \neq \pm 1$ we never meet
singular points, so there is no drop of the weight.

Notice that the special value $\lambda=-1$ brings in some drop of the
weights, with the sequence
 \begin{eqnarray}
\{ w_n \} = 1, 2, 4,8, 16, 30, 56, 104, 192,  \dots 
\end{eqnarray}
fitted by the rational generating function
\begin{eqnarray}
g(s) = \sum_k d_k s^k = \frac{1+s^4}{(1-s)(1-s-s^2-s^3)}
\end{eqnarray}
non vanishing entropy, no integrability.

\section{Folding}

We have shown, on a specific example, that a  differential-delay
equation, which was considered to have the other features of
integrability, also ``pass the algebraic entropy test''.

The key point allowing the calculation of the entropy was to consider
the values of the unknown function $u$ at different points $\dots,
t-1, t, t+1, \dots $ as members of a sequence of function verifying a
simple recurrence relation.

 Many  differential-delay equations will then pass the algebraic entropy
 test.
Starting from any differential-difference system of equations with
vanishing entropy we may turn it to a  differential-delay system with
zero entropy, that is to say a candidate to integrability.

Take for example the Ablowitz-Ladik system~\cite{AbLa76}:
\begin{eqnarray}
 \dot {q_n} = q_{n+1} - 2\; q_n + q_{n-1} + q_n \; r_n ( q_{n+1} +
 q_{n-1}) \label{abla1} \\ - \dot{r_n} = r_{n+1} - 2\; r_n + r_{n-1} +
 q_n \; r_n ( r_{n+1} + r_{n-1}).
\label{abla2}
\end{eqnarray}
This is an integrable semi-discretisation of the nonlinear
Schr\"odinger equation.  The time variable $t$ remained continuous, and
the space variable $x$ is discretised to $n$.  Folding the discrete
dimension $n$ over the continuous one $t $ leads to the
differential-delay system
\begin{eqnarray}
\dot{q}(t) &=& q(t+1) - 2\; q(t) + q(t-1) + 2\; q(t)\; r(t)\; ( q(t-1)
+q(t+1)) \label{folded1}\\ -\dot{r}(t) &=& r(t+1) - 2\; r(t) + r(t-1)
+ 2\; r(t)\; q(t) \; ( r(t-1) + r(t+1)) \label{folded2}
\end{eqnarray} 
which will pass the entropy test, since the system
(\ref{abla1},\ref{abla2}) does~\cite{DeVi12}, {\em because the
  calculations of both entropies are identical}.

{\bf Remark}: The case $a=0$ of equation (\ref{qcs}) [eq 18 of
  Ref~\cite{QuCaSa92})] can be obtained by this folding operation from
the original equation (4) of Ref.~\cite{Ma74}. This also means that
the latter has vanishing entropy as a semi-discrete equation.

Many more can be written using this rather rustic reduction, and of
course more sophisticated ones, in particular from the various symmetries
of integrable lattice equations for which the entropy has already been
shown to vanish in~\cite{DeVi12}.  We will get in this manner a number
of interesting autonomous and non-autonomous differential-delay
equations.

\section{Conclusion}

The vanishing of the algebraic entropy is a structural property of an
equation. The merit of the entropy test is that we may perform it
without knowing any of the solutions.  Solving the equations implies
to go beyond the discrete aspects, and to solve an interpolation
problem, which is hard, and usually requires some input on the nature
of the solution (domains of continuity, differentiability,
analyticity, ...).  The entropy test allows to choose a limited set of
equations of interest.


\end{document}